\documentclass[prb,aps,twocolumn,amsfonts,amssymb,amsmath,eqsecnum,floatfix]{revtex4}

\usepackage{bm}
\usepackage{graphicx}

\newcommand{\geklammert}[1]{\ensuremath{\left( #1\right)}}

\newcommand{\betrag}[1]{\ensuremath{\left| #1\right|}}
\newcommand{\eckklammern}[1]{\ensuremath{\left[ #1\right]}}
\newcommand{\formel}[1]{{Eq.~\geklammert{#1}}}	
\newcommand{\bild}[1]{{Fig.~#1}}	

\begin{document}
\title{Self-assembly of amphiphilic Janus particles at planar walls: A density functional study}
\author{Gerald~Rosenthal}
\email{rosenthal@itp.physik.tu-berlin.de}
\author{Sabine~H.~L. Klapp}
\email{klapp@physik.tu-berlin.de}
\affiliation{Institut f\"ur Theoretische Physik, Fakult\"at II, Technische Universit\"at Berlin, Sekr. EW 7-1,
Hardenbergstr. 36, D-10623 Berlin, Germany}

\begin{abstract}
We investigate the structure formation of amphiphilic molecules at planar walls using density functional theory.
The molecules are modeled as (hard) spheres composed of a hydrophilic and hydrophobic part. The orientation of the resulting Janus-particles is
described as a vector representing an internal degree of freedom. Our density functional approach involves Fundamental Measure Theory combined with
a mean-field approximation for the anisotropic interaction. Considering neutral, hydrophilic and hydrophobic walls, we study the
adsorption of the particles, focussing on the competition between the surface field and interaction-induced ordering phenomena. Finally, we consider
systems confined between two planar walls. It is shown that the anisotropic Janus interaction yields pronounced frustration effects at low temperatures.
\end{abstract}
\maketitle

\section{Introduction}
 Amphiphilic molecules are generally composed of a polar, hydrophilic (i.e. "water-loving") head and at least one hydrophobic (i.e., "fat-loving") tail, typically a carbon chain. The presence of these two ingredients, in combination with an aqueous solvent, yields
 a large variety of self-assembled structures on different length scales, ranging from molecular-size micelles to mesoscopic membranes, bicontinuous foams and lamellar phases. From an application point of view, amphiphiles are used in a variety of contexts, e.g., to reduce the surface tension in complex mixtures
(such as water and oil), as templates to create nanoporous materials \cite{Attard99} and, more recently, to stabilize bundle and network formation in solutions of carbon nanotubes \cite{Egorov,Angeliko08}.
 
From the theoretical side, the self-assembly of amphiphilic molecules has been investigated by a variety of approaches and models
\cite{schmid00,gompper94,shelley00}, including lattice gas systems \cite{bock04}, 
Ginzburg-Landau theory \cite{gompper94}, density functional studies of entropic models (where the surfactant is represented by a sphere plus an infinitely thin 
rod \cite{Ferber,Schmidt,Schmidt2}, or by a dimer \cite{Oxtoby}), and off-lattice simulations \cite{shinto00,goetz98} of flexible bead-spring molecules \cite{smit90}, with and without explicit solvent. 
A particularly simple, implicit-solvent model has been proposed by Tarazona {\em et al.} already in 1995 \cite{Tarazona}. Within this model, the amphiphilic particle is a sphere 
composed of two hemispheres, one being hydrophilic and the other one hydrophobic. The solvent-mediated interaction is then taken into account
via an effective directional potential involving the orientation vectors defined by the symmetry axis of the spheres.
Despite these simplifications (which completely neglect geometrical factors such as relative size of the head group etc.), the model
is capable of describing bilayer-, vesicle-, and micelle formation. From todays perspective, one would call this simplified amphiphilic molecules rather as "Janus spheres". This term was originally proposed by Casagrande {\em et al.} in 1989 \cite{Casagrande}
to describe spherical glass particles with one of the hemispheres being hydrophilic and the other one being hydrophobic. Since then, and as anticipated by P.~G. de Gennes on the occasion, of his Nobel lecture \cite{Gennes}, the surface properties of these Janus particles have become an area of great interest on their own. Generally, Janus particles are composed of at least two physically (or chemically) distinctive surfaces, and there has been significant progress to synthesize such particles \cite{Mueller,Granick,Carroll,Tsukruk}
(whose interactions can also be, e.g., dipolar or magnetic in character \cite{Duguet2, Duguet, Granick2, Kretzschmar}). Theoretically
there are several models "on the market" which are similar in character to that proposed by Tarazona {\em et al.} \cite{Tarazona}, but
have been independently proposed specifically for Janus particles \cite{Hess,Granick,telo}. Furthermore, in a recent study of Sciortino {\em et al.} \cite{Pastore}
Janus particles are described within a "patchy-particle" model.

In the present paper we are interested in the self-assembly of amphiphilic Janus particles at planar interfaces and in strongly confined situations
(slit-pores). Indeed, within the more general context of surfactant assembly, the impact of surfaces is important in a variety of applications, including synthesis of nanoporous materials, 
thin film deposition for lithographic processes and devices \cite{mitzi01}, and enhancement of chemical reactions \cite{haumann02}. An additional attractive factor 
of studying systems with surfaces is that the self-assembled structured can be experimentally studied
using e.g., neutron reflectometry (targeting the thickness of the layer) \cite{gilchrist99}, grazing incidence small-angle neutron scattering \cite{steitz04}
and atomic force microscopy (targeting the lateral structure) \cite{tiberg00}, and by studying adsorption isotherms \cite{Dietsch07}.
Moreover, there is a strong fundamental interest to understand the self-assembly process of Janus-like particles at interfaces 
as a bottom-up process for the design of future nanomaterials \cite{Millman05}.

Contrary to the extensively discussed case of true surfactant molecules at surfaces, the self-assembly of Janus particles at surfaces has so far rarely been considered, an exception
being a recent study of Hirose {\em et al.} \cite{Nonomura} who used a macroscopic theory based on the Young's equation.
In the present paper we use a microscopic approach, that is, classical density functional theory \cite{Evans,Henderson}. 
Following the original work of Tarazona \cite{Tarazona}, we employ a mean-field approximation to treat the anisotropic part of the interactions, whereas the repulsive (hard sphere) contribution is treated on a more sophisticated level. Specifically, we employ the so-called Fundamental Measure theory (FMT) \cite{Rosenfeld} which has turned out to be extremely successful for the description of inhomogeneous hard sphere systems \cite{Roth,Roth_FMT}. As a starting point, our investigations focus on planar structures such as bilayers. The key question is to determine thermodynamics conditions under which self-assembly arises at the surface, as compared to the corresponding bulk system. Moreover, we explore
the impact of different surface properties concerning, in particular, their hydrophilic or hydrophobic character.

The remainder of the paper is organized as follows. In Sec.~\ref{sec:Model} we introduce the model and also discuss its main features
as compared to other recently proposed models of Janus-like systems. The density functional formalism including the FMT contribution is presented  in 
Sec.~\ref{sec:Densityfunctionaltheory}, accompanied by an appendix which describes relevant technical issues. 
Numerical results are presented in Sec.~\ref{sec:Discussion}, where we focus on the case of single walls, but briefly discuss also the case
of Janus particles in slit-pore confinement.  We close the paper with a summary in Sec.~\ref{sec:Summary}.
\section{Theory}
\label{sec:Theory}
\subsection{Model}
\label{sec:Model}
 In this study we employ a simple, coarse-grained model of an amphiphilic system originally suggested by Tarazona and coworkers
 \cite{Tarazona}. In this model the amphiphilic molecules are represented by spherical particles consisting of two hemispheres, one being hydrophilic and the other one being hydrophobic. 
 Thus, the "molecules" rather resemble Janus particles. The solvent, which is omnipresent in real amphiphilic solutions, is treated implicitly. 
 The total pair potential between two Janus-like particles at positions $\mathbf{r}_1$ and $\mathbf{r}_2$ subdivides into a hard-sphere (HS) and an anisotropic contribution,
  \begin{equation}
    \phi\geklammert{\mathbf{r}_{12},\hat{\mathbf{u}}_1,\hat{\mathbf{u}}_2}=\phi^{\mathrm{HS}}\geklammert{r_{12}}+\phi^{\mathrm{I}}\geklammert{\mathbf{r}_{12},\hat{\mathbf{u}}_1,\hat{\mathbf{u}}_2},\label{potential}
  \end{equation}
 where $\mathbf{r}_{12}=\mathbf{r}_1-\mathbf{r}_2$, and the HS potential is given by
  \begin{equation}  
    \phi^{\mathrm{HS}}\geklammert{r_{12}}=\begin{cases}
					    \infty,& r_{12}<\sigma,\\
					    0,& r_{12}>\sigma,
					  \end{cases}
  \end{equation}
  with $r_{12}=\betrag{\mathbf{r}_{12}}$ and $\sigma$ being the HS diameter. Further, $\hat{\mathbf{u}}_1$ and $\hat{\mathbf{u}}_2$ are unit
  vectors denoting the orientation of the sphere. Specifically, $\hat{\mathbf{u}}_i$ ($i=1,2$) points from the hydrophobic to the hydrophilic side of particle $i$. 
  The vectors $\hat{\mathbf{u}}=\hat{\mathbf{u}}\geklammert{\bm{\omega}}$ are parameterized 
  by $\bm{\omega}=\geklammert{\varphi,\theta}$, with $\varphi = [0,2\pi]$ and $\theta = [0,\pi]$.
  The anisotropic interaction is defined as
  \begin{equation}  
    \phi^{\mathrm{I}}\geklammert{\mathbf{r}_{12},\hat{\mathbf{u}}_{1},\hat{\mathbf{u}}_{2}}=\phi_{1}\geklammert{r_{12}}\geklammert{\hat{\mathbf{u}}_1 - \hat{\mathbf{u}}_2}\cdot
    \hat{\mathbf{r}}_{21},\label{janus_potential}
  \end{equation}
  where
  \begin{equation} 
    \phi_{1}\geklammert{r_{12}}=\begin{cases}
			0,& r_{12}<\sigma,\\
			C\,\mathrm{exp}\left(-\lambda\left(r_{12}-\sigma\right)\right)/r_{12},& r_{12}>\sigma.
		\end{cases}\label{yukawa_pot}
  \end{equation}
  In \formel{\ref{yukawa_pot}} the parameters $C$ and $\lambda$ measure the coupling strength and the (inverse) range of the interaction, respectively. If not stated otherwise,
  we set $\lambda\sigma=3$.
  
  A characteristic feature of the anisotropic potential $\phi^{\mathrm{I}}$ is that it lacks of any coupling between the orientations $\hat{\mathbf{u}}_i$ of the two involved particles ($i=1,2$).
  In fact, \formel{\ref{janus_potential}} represents only the lowest-order anisotropic contribution of a more general expansion of the pair potential between linear
  molecules into spherical harmonics \cite{Gubbins} (see also Eq.~(\ref{janus_potential2}) below).\\
  In \bild{\ref{Model}} we sketch the most important configurations of two particles; the corresponding pair energies according to the angle-dependent terms in Eq.~(\ref{janus_potential})
  are given in the related table (first column). 
  \begin{figure}
\includegraphics[width=7cm,angle=-90]{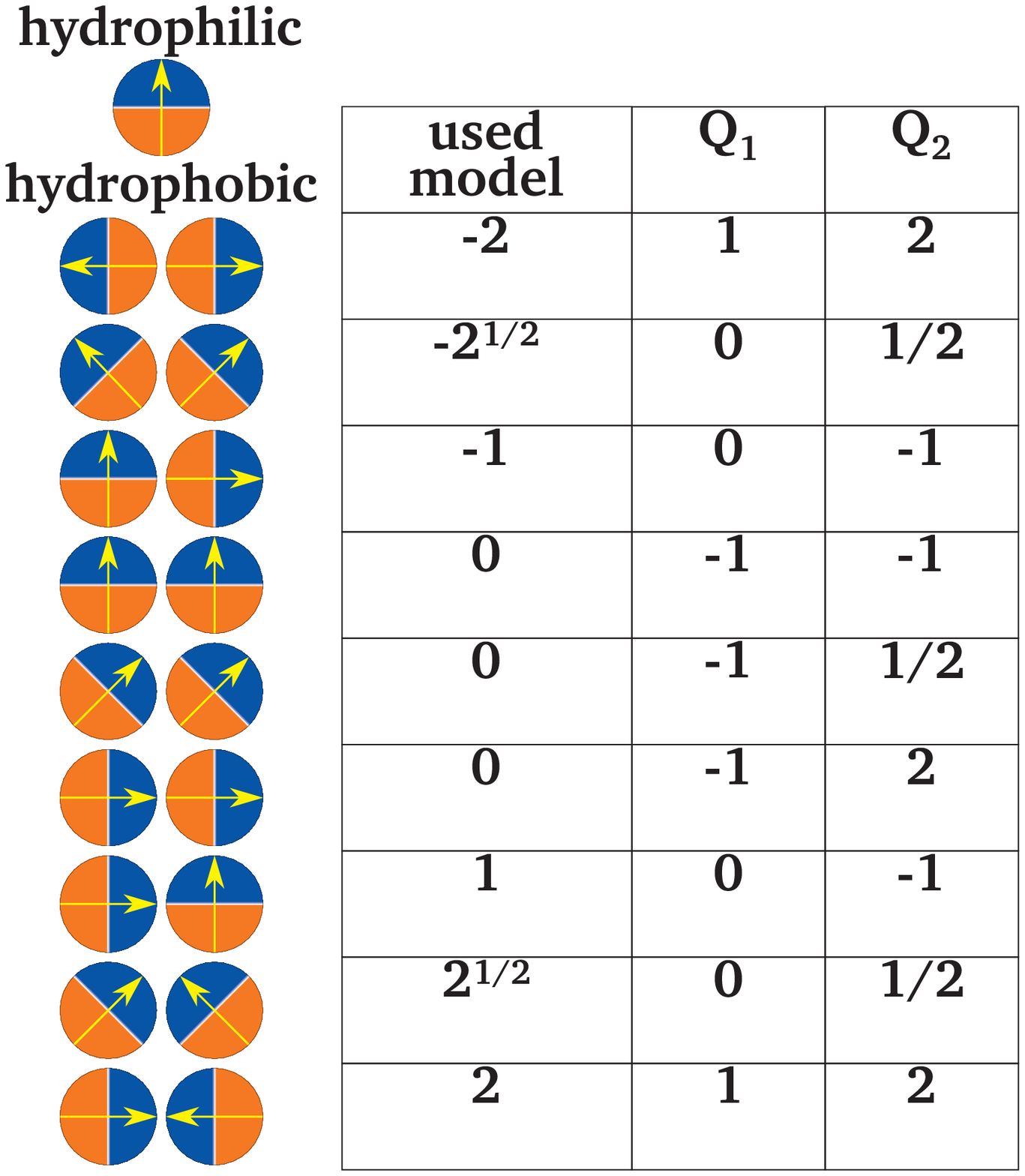}
\caption{Left: Illustration of typical pair configurations within our model. Right: the first column of the table gives the orientational contributions to the pair energies [i.e., the term
$({\hat{\mathbf{u}}_1 - \hat{\mathbf{u}}_2})\cdot
    \hat{\mathbf{r}}_{21}$]
according to \formel{\ref{janus_potential}}. The second and third columns refer to the possible extensions of the model according to Eqs.~(\ref{q1_add_pot}) and 
 (\ref{q2_add_pot}), respectively. \label{Model}} 
\end{figure}
Within our model the particles prefer to be orientated in opposite directions, such that the hydrophobic sides point towards one another. The opposite configuration
(with facing hydrophilic sides) is the energetically most unfavorable one. Further, parallel orientations are energetically neutral. The strong energetic preference (or penalty) of
configurations with facing hydrophobic (hydrophilic) sides mimics the effects expected in a real system which would include water as a solvent. Clearly, the 
 water molecules will preferentially adsorb at the hydrophilic side of each particle. The resulting steric exclusion yields an effective repulsion of the hydrophilic sides of neighboring 
 Janus particles. On the other hand, the fact that the hydrophobic sides dislike water effectively favors configurations where these sides point towards each other, such that the
 contact with water is minimized. An additional source of attraction between hydrophobic sides can arise from a functionalization with, e.g., Thiol. An example has been described in 
 Ref.~\onlinecite{Granick2}
 where Thiol (and other molecular groups) is used to functionalize the gold-coated side of a polystyrene sphere. The functionalization yields a depletion area with respect to water
 and thus, an additional effective attraction between the hydrophobic (gold) hemispheres.

It is interesting to briefly consider the subsequent terms [i.e., beyond Eq.~(\ref{janus_potential})] in the general expansion of the anisotropic pair potential. These terms read
\cite{Tarazona}
  \begin{multline}  
    \phi^{\mathrm{I}}\geklammert{\mathbf{r}_{12},\hat{\mathbf{u}}_{1},\hat{\mathbf{u}}_{2}}=\phi_{1}\geklammert{r_{12}}\geklammert{\hat{\mathbf{u}}_1 - \hat{\mathbf{u}}_2}\cdot
    \hat{\mathbf{r}}_{21}\\
    +q_1\phi_2\geklammert{r_{12}}Q_1\geklammert{\hat{\mathbf{r}}_{21},\hat{\mathbf{u}}_{1},\hat{\mathbf{u}}_{2}}
    +q_2\phi_3\geklammert{r_{12}}Q_2\geklammert{\hat{\mathbf{r}}_{21},\hat{\mathbf{u}}_{1},\hat{\mathbf{u}}_{2}}\label{janus_potential2},
  \end{multline}
  where
  \begin{multline} 
    Q_1\geklammert{\hat{\mathbf{r}}_{21},\hat{\mathbf{u}}_{1},\hat{\mathbf{u}}_{2}}=\biggl{(}-\geklammert{\hat{\mathbf{u}}_1\cdot\hat{\mathbf{r}}_{21}}
    \geklammert{\hat{\mathbf{u}}_2\cdot\hat{\mathbf{r}}_{21}}\\
-\sqrt{1-\geklammert{\hat{\mathbf{u}}_1\cdot\hat{\mathbf{r}}_{21}}^2}\sqrt{1-\geklammert{\hat{\mathbf{u}}_2\cdot\hat{\mathbf{r}}_{21}}^2}\mathrm{cos}\geklammert{\varphi_1-\varphi_2}\biggr{)}\label{q1_add_pot}\\\\
=-\hat{\mathbf{u}}_{1}\cdot\hat{\mathbf{u}}_{2},
  \end{multline}
  and
  \begin{equation} 
    Q_2\geklammert{\hat{\mathbf{r}}_{21},\hat{\mathbf{u}}_{1},\hat{\mathbf{u}}_{2}}=\frac{1}{2}\geklammert{3\geklammert{\hat{\mathbf{u}}_1\cdot\hat{\mathbf{r}}_{21}}^2
    +3\geklammert{\hat{\mathbf{u}}_2\cdot\hat{\mathbf{r}}_{21}}^2-2}.\label{q2_add_pot}
  \end{equation}
  In Eq.~(\ref{janus_potential2}), $\phi_2\geklammert{r_{12}}$ and $\phi_3\geklammert{r_{12}}$ are distance-dependent functions that could be set, for example, equal to the Yukawa-like function
  $\phi_1\geklammert{r_{12}}$ defined in Eq.~(\ref{yukawa_pot}).
  The implications of the additional, angle-dependent contributions are illustrated in the center and the right column of the table in \bild{\ref{Model}}. Equation~(\ref{q1_add_pot}) 
  adds a term ($Q_1$) involving the negative scalar product of the orientation vectors; this term clearly prefers parallel orientations
  irrespective of the configuration of the particles relative to the connecting vector.
  The term $Q_2$ defined in
  \formel{\ref{q2_add_pot}} favors parallel orientations
  perpendicular to the connecting axis. 
  Here we focus on the lowest-order term [cf.~Eq.~(\ref{janus_potential})] which describes the main characteristics of our amphiphilic system \cite{Tarazona}.
    
  To complete the discussion on the interaction between the Janus particles we briefly compare 
 the present model with other models that were recently proposed in the literature. 
 The potential suggested by Hess~\textit{et al.} \cite{Hess} involves all types of angle-dependent terms appearing in Eqs.~(\ref{janus_potential})-(\ref{q2_add_pot}). However, the parameters
 in Ref.~\onlinecite{Hess} were chosen such that {\em parallel} orientation of the Janus spheres is preferred, contrary to our model favoring {\em antiparallel} alignment
 (see Fig.~\ref{Model}). 
 Another Janus potential has been suggested by Sciortino {\em et al.}  \cite{Pastore}.
 Within this model, the (spherical) particles have one patch; the patches of neighboring particles then interact via a square-well potential. The remaining 
 parts of the spheres only induce steric repulsion. This picture is quite different from our model, which 
 should rather be compared to a particle with two patches, one mimicking the hydrophilic and the other one the hydrophobic part. Moreover, 
 in our model the strongest attractive (repulsive) interactions occur when
  the particles are aligned in an antiparallel (parallel) manner. The one-patch model in Ref.~\onlinecite{Pastore} concentrates on the attraction in "antiparallel" configurations,
 i.e. configurations where the two patches become coupled through the square-well zone.
  
  Our main goal in the present paper is to study the amphiphilic system in the presence of a planar surface
  (located at $z=0$). The simplest surface model is a hard wall, yielding the surface potential
  \begin{equation}
    \phi^{\mathrm{surf}}\geklammert{z}=\begin{cases}
					    \infty,& z<\sigma/2,\\
					    0,& z>\sigma/2.
					  \end{cases}
					  \label{surf_hs}
  \end{equation}  
  To include effects of preferential adsorption we also consider potentials of the form
  \begin{equation}
    \phi^{\mathrm{surf}}\geklammert{z,\hat{\mathbf{u}}}=\begin{cases}
      \infty,& z<\sigma/2,\\
	  C_2\,\mathrm{exp}\geklammert{-\lambda_2\geklammert{z-\sigma/2}}\hat{\mathbf{u}}\cdot\hat{\mathbf{e}}_z/z,& z>\sigma/2,
      \end{cases}
      \label{surf_pref}
  \end{equation}
  where the unit vector $\hat{\mathbf{e}}_z$ points along the z-direction.
  The sign of the parameter $C_2$ determines the preferred orientations of the particles. Specifically, for negative (positive) values of $C_2$ the surface prefers the hydrophobic (hydrophilic)
  side of the Janus particle.
  Finally, the range of the surface potential is controlled by the parameter $\lambda_2$. For simplicity, we set $\lambda_2=\lambda=3\sigma^{-1}$.
  %
\subsection{Density functional theory}
\label{sec:Densityfunctionaltheory}
  The central quantity in our study is the normalized singlet density
  \begin{equation}
   \rho\geklammert{\mathbf{r},\bm{\omega}}=\rho\geklammert{\mathbf{r}}\alpha\geklammert{\mathbf{r},\bm{\omega}},\int d^2\omega\alpha\geklammert{\mathbf{r},\bm{\omega}}=1,
  \end{equation}
  which depends on both, the position and the orientation of the particles. In the equation above, $d^2\omega = d\varphi d\theta\mathrm{sin}\theta$.
  Following other DFT studies of molecular fluids (see, e.g., Refs.~\cite{Tarazona,Gramzow,Range}) we write the singlet density as a product of a position-dependent number density $\rho\geklammert{\mathbf{r}}$
  and an orientational distribution function $\alpha\geklammert{\mathbf{r},\bm{\omega}}$.
  The equilibrium singlet density is obtained from the grand-canonical density functional $\Omega\eckklammern{\rho}$, which is given as
    \begin{multline}
    \Omega\left[\rho,\alpha\right]=\beta^{-1}\int d^{3}r\rho\geklammert{\mathbf{r}}\geklammert{\mathrm{ln}\geklammert{{\rho\geklammert{\mathbf{r}}\Lambda^3}}-1}\\
    +\beta^{-1}\iint d^{3}r d^{2}\omega\rho\geklammert{\mathbf{r}}\alpha\geklammert{\mathbf{r},\bm{\omega}}\mathrm{ln}\geklammert{4\pi\alpha\geklammert{\mathbf{r},\bm{\omega}}}\\
    +\iint d^{3}r d^{2}\omega\rho\geklammert{\mathbf{r}}\alpha\geklammert{\mathbf{r},\bm{\omega}}\phi^{\mathrm{surf}}\geklammert{\mathbf{r},\hat{\mathbf{u}}\geklammert{{\bm{\omega}}}}\\
    -\mu\int d^{3}r\rho\geklammert{\mathbf{r}}+\beta^{-1}\int d^{3}r\phi^{\mathrm{WBII}}\geklammert{\left\{ n_{\alpha}\geklammert{\mathbf{r}}\right\} }\\
    +\iiint d^{3}rd^{3}r'd^{2}\omega\rho\geklammert{\mathbf{r}}\rho\geklammert{\mathbf{r'}}\\
    \cdot\alpha\geklammert{\mathbf{r},\bm{\omega}}\phi_{1}\geklammert{\mathbf{r}-\mathbf{r'}}\frac{\mathbf{r'}-\mathbf{r}}{\left(\mathbf{r}-\mathbf{r'}\right)}\cdot\hat{\mathbf{u}}
    \geklammert{{\bm{\omega}}}.
    \label{omega}
  \end{multline}
 The first two contributions on the right side of Eq.~(\ref{omega}) represent the ideal gas part of the free energy, with $\beta=1/k_\mathrm{B}T$ (where $k_\mathrm{B}$ is the Boltzmann constant,
  $T$ the temperature)  and $\Lambda$ is the thermal de Broglie wavelength.
  Further, the third and the fourth term include contributions from the external (surface) potential and the chemical potential, respectively.
  The fifth term arises from the HS interactions, which we treat by FMT \cite{Rosenfeld} as described in more detail below. Finally, 
  the last term in Eq.~(\ref{omega}) involves
  the anisotropic interactions~(\ref{janus_potential}). We treat this free-energy contribution by a mean-field approximation (which can be rationalized from a
 $\mathrm{\lambda}$-expansion\cite{HansenMcDonald} of the pair potential). In this framework, the pair correlation function is set to one for separations $r_{12}>\sigma$.
 Thus, correlation effects beyond the core condition are neglected within the anisotropic contribution to the density functional.
  
 Regarding the HS free energy, we use the White-Bear mark II version\cite{Roth} of FMT. In this framework the function $\phi^{\mathrm{WBII}}$ appearing in the fifth term on the right side of \formel{\ref{omega}} is given by
  \begin{multline}
    \phi^{\mathrm{WBII}}\geklammert{\left\{ n_{\alpha}\geklammert{\mathbf{r}}\right\} }=-n_0\geklammert{\mathbf{r}}\mathrm{ln}\geklammert{1-n_3\geklammert{\mathbf{r}}}\\
    +\frac{n_1\geklammert{\mathbf{r}}n_2\geklammert{\mathbf{r}}-\mathbf{n}_1\geklammert{\mathbf{r}}\cdot\mathbf{n}_2\geklammert{\mathbf{r}}}{1-n_3\geklammert{\mathbf{r}}}\geklammert{1+\frac{\psi_2\geklammert{n_3\geklammert{\mathbf{r}}}}{3}}\\
      +\frac{{n_2}^3\geklammert{\mathbf{r}}-3n_2\geklammert{\mathbf{r}}{\mathbf{n}_2}^2\geklammert{\mathbf{r}}}{24\pi\geklammert{1-n_3\geklammert{\mathbf{r}}}^2}\geklammert{1-\frac{\psi_3\geklammert{n_3\geklammert{\mathbf{r}}}}{3}}.\label{FMT_1}
  \end{multline}  
  In \formel{\ref{FMT_1}},
  \begin{multline}
    \psi_2\geklammert{n_3\geklammert{\mathbf{r}}} =\frac{1}{n_3\geklammert{\mathbf{r}}}{\big(}2n_3\geklammert{\mathbf{r}}-{n_3}^2\geklammert{\mathbf{r}}\noindent\\
      +2\geklammert{1-n_3\geklammert{\mathbf{r}}}\mathrm{ln}\geklammert{1-n_3\geklammert{\mathbf{r}}}{\big)},
  \end{multline} 
  and
  \begin{multline}
    \psi_3\geklammert{n_3\geklammert{\mathbf{r}}} =\frac{1}{{n_3}^2\geklammert{\mathbf{r}}}{\big(}2n_3\geklammert{\mathbf{r}}-3{n_3}^2\geklammert{\mathbf{r}}+2{n_3}^3\geklammert{\mathbf{r}}\\
    +2\geklammert{1-n_3\geklammert{\mathbf{r}}}^2\mathrm{ln}\geklammert{1-n_3\geklammert{\mathbf{r}}}{\big)}.
  \end{multline}
  In the expressions above, the scalar functions $n_\alpha\geklammert{\mathbf{r}}$ with $\alpha = 0,1,2,3$ and the vector functions $\mathbf{n}_\gamma\geklammert{\mathbf{r}}$ with $\gamma=1,2$ are weighted densities defined as
  \begin{equation}
    {n_\alpha\geklammert{\mathbf{r}} \choose \mathbf{n}_\alpha\geklammert{\mathbf{r}}}=\int d^3 r' \rho\geklammert{\mathbf{r}-\mathbf{r'}}{\bar\omega_\alpha\geklammert{\mathbf{r}'} \choose \bar{\bm{\omega}}_\alpha\geklammert{\mathbf{r}'}}.\label{weighteddensities}
  \end{equation}
  The weight functions are composed of Dirac delta distributions $\delta(x)$ and Heaviside step functions $\Theta(x)$. Explicitely, one has 
  \begin{equation}
	\bar{\omega}_3\geklammert{\mathbf{r}}=\Theta\geklammert{R - \betrag{\mathbf{r}}},\bar{\omega}_2\geklammert{\mathbf{r}}=\delta\geklammert{R - \betrag{\mathbf{r}}},
  \end{equation}
  \begin{equation}
	\bar{\omega}_1\geklammert{\mathbf{r}}=\frac{\bar{\omega}_2\geklammert{\mathbf{r}}}{4\pi R},\bar{\omega}_0\geklammert{\mathbf{r}}=\frac{\bar{\omega}_2\geklammert{\mathbf{r}}}{4\pi R^2},
  \end{equation}
  \begin{equation}
	\bar{\bm{\omega}}_2\geklammert{\mathbf{r}}=\delta\geklammert{R - \betrag{\mathbf{r}}}\frac{\mathbf{r}}{\betrag{r}},\bar{\bm{\omega}}_1\geklammert{\mathbf{r}}=\frac{\bm{\bar{\omega}}_2\geklammert{\mathbf{r}}}{4\pi R},
  \end{equation}
  where $R=\sigma/2$.
  
  The equilibrium density is found by minimization \cite{Evans} of the grand canonical functional given in \formel{\ref{omega}}, that is
  \begin{equation}
      \frac{\delta\Omega\eckklammern{\rho,\alpha}}{\delta\rho\geklammert{\mathbf{r}}}{\Big|}_{\rho_0\geklammert{\mathbf{r}}}=0,
  \end{equation}
  \begin{equation}
      \frac{\delta\Omega\eckklammern{\rho,\alpha}}{\delta\alpha\geklammert{\mathbf{r},\bm{\omega}}}{\Big|}_{\alpha_0\geklammert{\mathbf{r},\bm{\omega}}}=0.
  \end{equation}
  Due to the mean-field approximation of the anisotropic interaction in the density functional [see last term in Eq.~(\ref{omega})], combined with the fact that this interaction
  is linear in the orientations [see \formel{\ref{janus_potential}}],
  minimization with respect to $\alpha\geklammert{\mathbf{r},\bm{\omega}}$  yields the simple, explicit expression
  \begin{equation}
    \alpha_0\geklammert{\mathbf{r},\bm{\omega}}=\frac{\mathrm{exp}\geklammert{\mathbf{a}\geklammert{\mathbf{r}}\cdot\hat{\mathbf{u}}\geklammert{\bm{\omega}}}}{\int d^{2}\omega' \mathrm{exp}\geklammert{\mathbf{a}\geklammert{\mathbf{r}}\cdot\hat{\mathbf{u}}\geklammert{\bm{\omega}'}}},
  \end{equation}
  where
  \begin{equation}
    \mathbf{a}\geklammert{\mathbf{r}_1}=\beta\int d^{3}r_2\rho\geklammert{\mathbf{r}_2}\phi_1\geklammert{\mathbf{r}_{12}}\hat{\mathbf{r}}_{12}.
  \end{equation}

  Contrary to the orientational distribution, the position-dependent number density has to be calculated numerically. Technical details are given in the Appendix, where we also
  give the exact boundary conditions by which the numerical results can be checked.
  To restrict the numerical effort we focus on density distributions that depend only on $z$. Clearly, this strategy implies the possibility of missing other, and maybe
  energetically preferable solutions. Indeed, by considering both, $z$-dependent and $r$-dependent density distributions, 
 Tarazona~\textit{et al.} \cite{Tarazona} showed that 
 the present model yields planar structures (''membranes"), but also spherical structures ("vesicle" and "micelles") under bulk conditions.
 However, in the presence of surfaces, a full three-dimensional calculation would demand considerable computational cost.
  Therefore, we focus on density distributions with the symmetry suggested by the planar wall and assume translational symmetry in the other spatial dimensions. 
  As a consequence, the orientational distribution also simplifies to $\alpha\geklammert{\mathbf{r},\bm{\omega}}=\alpha(z,\theta)$.
    
  Having this in mind, it is useful to introduce various $z$-dependent orientational order parameters. Specifically, we define  
  \begin{equation}
    h\geklammert{z}=2\pi\int_0^{\pi} d\theta\,\mathrm{sin}\theta\,\alpha\geklammert{z,\theta}\mathrm{cos}\theta
    \label{hz}
  \end{equation}
 and 
  \begin{equation}
    h_2\geklammert{z}=\pi\int_0^{\pi} d\theta\,\mathrm{sin}\theta\,\alpha\geklammert{z,\theta}\geklammert{3\mathrm{cos}^2\theta-1}.
    \label{h2z}
  \end{equation}
  The function $h(z)$ describes the "polarization". It can take values between $-1$ and $1$ corresponding to complete alignment in negative (positive) $z$-direction at the position $z$ considered.
  The latter, second-rank order parameter $h_2(z)$ also describes the ordering along the $z$-axis, but it is independent of the direction. Within our (mean-field) approximation, the function
  $h_2(z)$ is strongly coupled to $h(z)$, that is, $h\geklammert{z}\neq 0$ automatically yields 
  $h_2\geklammert{z}\neq 0$ and vice versa (see also \bild{\ref{Eta03}}). Hence, we mainly consider the $z$-dependent "polarization", $h\geklammert{z}$.
%
\section{Results and discussion}
\label{sec:Discussion}
\subsection{Bulk behavior}
\label{bulk_system}
\begin{figure}
\includegraphics[width=8cm]{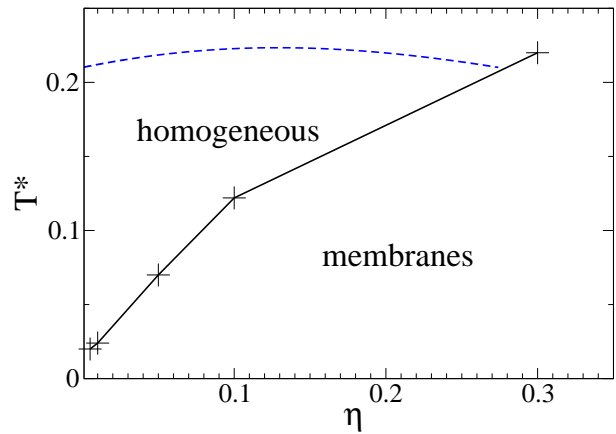}
\caption{Bulk phase diagram. The solid line indicates the estimated transition temperatures (see main text) between the homogeneous, isotropic high-temperature phase and an 
ordered low-temperature phase characterized by membranes. The dashed line corresponds to the vapor-liquid coexistence curve resulting from a density function calculation with a MMFT treatment of the anisotropic interactions (yet under the assumption of a homogeneous isotropic state) \label{bulk}.} 
\end{figure}

As a starting point we briefly consider the bulk system of Janus spheres, which may be characterized by the packing fraction, $\eta=(\pi/6)\rho\sigma^3$, and the reduced
temperature $T^{*}=k_{\mathrm{B}}T/C$ [with $C$ being the coupling constant appearing in Eq.~(\ref{yukawa_pot})]. With the latter definition, the limit $T^{*}\rightarrow\infty$ corresponds to a pure HS system. Therefore, at high $T^{*}$ and densities $\eta\lesssim 0.45$, minimization of the density functional~(\ref{omega}) just yields a homogeneous, isotropic phase characterized
by constant number density, $\rho(z)=\rho$, and constant orientational distribution, $\alpha(z,\theta)=1/4\pi$. Upon lowering the temperature, inhomogeneous, ''membrane''-like solutions
can appear. By ''membrane" we refer to an isolated bilayer of Janus particles, which is infinitely extended along the $x$- and $y$-directions, and 
where the two sub-layers are oriented towards each other in the energetically preferable configuration
(with facing hydrophobic sides). The appearance of such membranes was already shown in the original paper of Tarazona {\em et al.} \cite{Tarazona}. 
To detect the corresponding temperature range in the present study, we start the numerical minimization 
with an initial
guess for $\rho(z)$ composed of two Gaussians. At high temperatures, this initial profile quickly disappears during the iteration procedure.
For low values of $T^{*}$, on the other hand, the procedure yields indeed formation of periodically repeated bilayer structures, characterized by a sharp decay of the density
outside of the bilayers, and a depletion zone between the membranes. 
The temperatures separating both regions are indicated by the solid line in Fig.~\ref{bulk}. 
However, the precise location of this line has to be considered with some caution since the inhomogeneous low-temperature profile did not converge to a truly stable result. Therefore, we should consider the line in Fig.~\ref{bulk} rather as an upper limit of the temperature
range related to membrane formation. We also note that our results differ from those obtained in Ref.~\onlinecite{Tarazona}, where the transition temperatures were found to be higher (the focus in \cite{Tarazona} was on lower densities, though). These deviations might be due
to the fact that the HS part of the density functional used in Ref.~\onlinecite{Tarazona} was treated in weighted-density approximation, rather than by FMT as in the present study.

We also note that the present functional does not predict a gas-liquid transition of the Janus-particle system, as it was found, e.g., in a recent simulation study of a related
model system consisting of patchy Janus particles \cite{Pastore}. Within our study, the absence of condensation is a consequence of the mean-field treatment of the anisotropic interactions; indeed,
the corresponding term in the density functional [see last term on the right side of Eq.~(\ref{omega})] is zero in a fully disordered system. Interestingly, if we focus 
on such a homogeneous, isotropic system and use a {\em modified} mean field approximation (MMFT) for the anisotropic part (for applications of the MMFT to related
systems, see Refs.~\cite{Gramzow,Range}), 
we do find a gas-liquid phase transition, indicating that the Janus interaction is {\em effectively attractive}.
The corresponding phase coexistence points are indicated by the dashed line in Fig.~\ref{bulk}. Here we do not explore this aspect further, 
the main reason being that the use of MMFT in the presence of surfaces (which are the focus of our study) would have implied a drastic increase of numerical effort.
In the following discussion of surface effects (on the basis of the pure mean-field approximation)
we therefore consider temperatures and packing fractions in the whole range above the solid line in Fig.~\ref{bulk}, i.e., where the bulk system is disordered.
\subsection{Surface effects at neutral walls}
\label{neutral_walls}
\begin{figure}
\includegraphics[width=8cm]{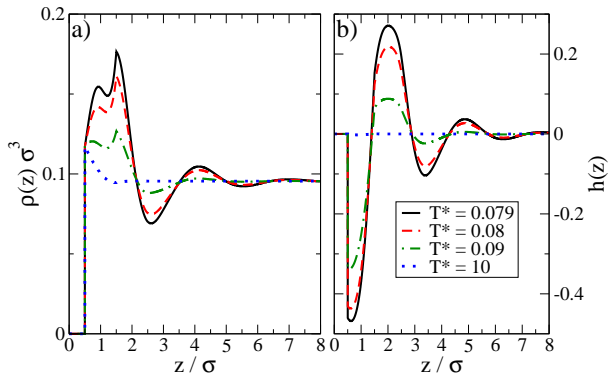}
\caption{(a) Density profiles and (b) order parameters at $\eta=0.05$ and various temperatures. Here and in the subsequent figures, the hard wall is located at $z=0$,
and the range parameter is set to $\lambda\sigma=3$.\label{densities005}} 
\end{figure}

 We start the discussion of surface effects with the case of hard ("neutral") walls [see Eq.~(\ref{surf_hs})].
 In \bild{\ref{densities005}}(a) we present density profiles for a dilute system characterized by $\eta=0.05$ and different temperatures $T^{*}\geq T^{*}_{\mathrm{m}}$,
 where $T^{*}_{\mathrm{m}}\approx 0.079$ is the temperature related to membrane formation in the bulk (see Fig.~\ref{bulk}). 
The corresponding order parameter function $h(z)$ [see Eq.~(\ref{hz})] is plotted in Fig.~\ref{densities005}(b). At $T^{*}=10$,  the density profile agrees within the numerical
 accuracy with that of a HS fluid, consistent with the negligible values of the order parameter. By decreasing the temperature, 
 configurations with neighboring hydrophobic sides of the particles become more and more favorable, yielding orientational ordering of the particles as visible in Fig.~\ref{densities005}(b). 
 At the same time, the density profiles change from the typical behavior of a dilute HS system, where the density maximum occurs directly at the wall, towards a soft, loosely packed structure 
 where most of the particles agglomerate somewhat away from the wall [see, e.g., data at $T^{*}=0.08$ in Fig.~\ref{densities005}(a)].
 Moreover, the distance between the first two layers indicated
 by the two close maxima in $\rho(z)$ is smaller than one particle diameter. Analyzing the corresponding orientational order parameters we find that, 
 in the first layer (located at $z\approx 1\sigma$) the particles are oriented such that their hydrophilic side points preferentially to the wall, whereas in the second layer (at $z\approx 1.8-2\sigma$), 
 they tend to orient in the opposite way.  Thus, despite the low density considered, a significant degree of orientational ordering is already present.
  
\begin{figure}
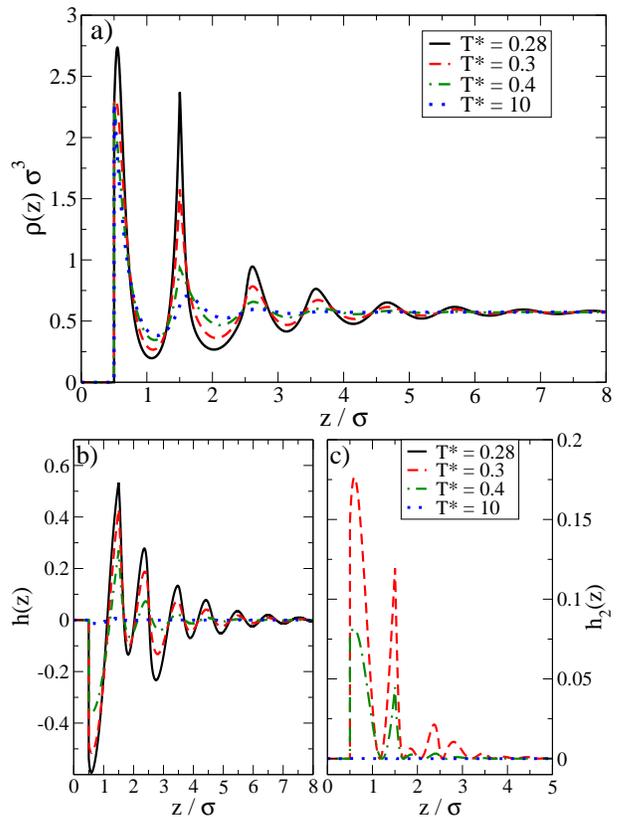

\includegraphics[width=8cm]{Fig4a.eps}
\includegraphics[width=8cm]{Fig4bc.eps}
\caption{Density profiles (a) and order parameter functions $h(z)$ (b) and $h_2(z)$ (c) at $\eta=0.3$ and various temperatures.\label{Eta03}} 
\end{figure}

 As expected, both the translational and the orientational ordering becomes more pronounced upon an increase of the density. This is illustrated
 in Fig.~\ref{Eta03} where we plot density and orientational profiles for the exemplary case $\eta=0.3$.
At the lowest temperature considered, $T^{*}=0.28$, the Janus particles close to the wall are almost completely 
arranged into a double layer, as reflected by the high values of the first, pronounced maxima in $\rho(z)$ located at $z=0.5\sigma$ and $z=1.5\sigma$. The double-layer formation
is accompanied by a high degree of ''polar'' orientational order, as seen from the pronounced change of the function $h(z)$ 
[see Fig.~\ref{Eta03}(b)] 
from negative values at $z\approx 0.5\sigma$
(indicating that the hydrophilic side tend to point towards the wall) to positive values at $z\approx 1.5\sigma$. Part~(c) of Fig.~\ref{Eta03} 
additionally shows the function $h_2(z)$ defined in Eq.~(\ref{h2z}); however, as anticipated in Sec.~\ref{sec:Densityfunctionaltheory}, this function just follows the oscillations of 
$h(z)$ and will thus not be considered further. Combing back to the density profile $\rho(z)$ [see Fig.~\ref{Eta03}(a)]
we see that, at distances $z$ beyond the bilayer at the wall, 
the third maximum appears only at $z\approx 2.7\sigma$, indicating that the next layer is slightly shifted towards
larger separations. This is a result of the repulsion between the hydrophilic sides of a particle in the second, and one in the third layer. Indeed, the preferred orientation of the third layer
becomes apparent from the positive value of $h(z)$ in Fig.~\ref{Eta03}(b). We stress that all of these orientational ordering effects occur at temperatures where the corresponding
bulk system is still homogeneous (see Fig.~\ref{bulk}).

Within the framework of neutral walls, we have also considered the influence of the range parameter $\lambda$ within the Janus-particle
interaction [see Eq.~(\ref{yukawa_pot})]. It turns out, however, that the latter has no crucial effect
for the systems considered here. Upon increase of $\lambda$ (i.e., decrease of the interaction range) one merely observes a decrease of the density maxima close to the walls, and a damping of the
oscillations at larger distances.
%
 \subsection{Influence of the surface fields}
 \label{surface_fields}
\begin{figure}
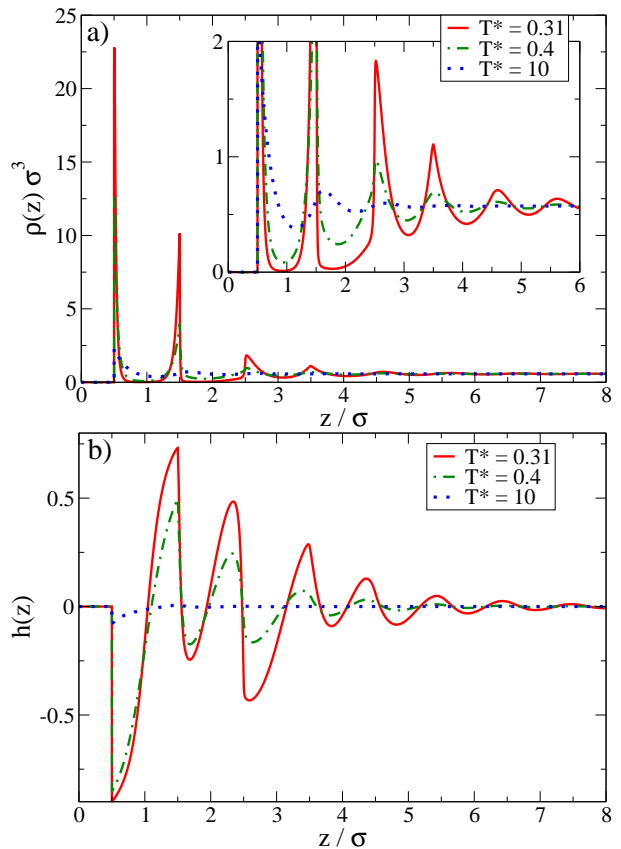

\includegraphics[width=8cm]{Fig5a.eps}
\includegraphics[width=8cm]{Fig5b.eps}
\caption{(a) Density profiles and (b) order parameters at $\eta=0.3$ and a hydrophilic wall ($C_2=C$, $\lambda_2\sigma= \lambda\sigma=3$). The inset in part a)
shows the density profile on an enlarged scale. \label{hydrophilic}} 
\end{figure}

We next consider the ordering behavior in systems, where the pure confinement effect induced by a neutral wall is supplemented by surface fields 
preferring (locally) a specific orientation of the Janus particles [see Eq.~(\ref{surf_pref})]. We start with a wall preferring the hydrophilic sides (such as  silica
\cite{Dietsch07}). This situation implies
that the surface field {\em supports} the orientation already found close to neutral walls (see Sec.~\ref{neutral_walls}).
Exemplary density profiles and order parameters are shown in Fig.~\ref{hydrophilic}, where the bulk density $\eta=0.3$, and
the fluid-fluid and fluid-surface interactions (as well as the corresponding range parameters) are of the same magnitude.
Due to the hydrophilic (and thus, supportive) character of the wall, the density in the contact layer, as well as the corresponding orientational order parameter, is even enhanced as compared 
to the corresponding system with neutral walls. This can be seen when we compare, e.g., results for the temperature $T^{*}=0.31$ in Figs.~\ref{hydrophilic}(a) and (b)
with corresponding ones at $T^{*}=0.3$ in Fig.~\ref{Eta03}. Contrary the neutral-wall case, however, the second density maximum 
at the hydrophilic wall is much smaller than the first. We understand this behavior as a consequence of the fact that, in the second layer, the hydrophilic orientation preferred by the wall competes with that dictated by the fluid-fluid-interaction; the latter rather favors the hydrophobic side orienting towards the wall. The competition is also reflected by the asymmetric shape
of the maximum in the function $h(z)$. Finally, considering the third density maximum and comparing with Fig.~\ref{Eta03}, we find that this (and the subsequent) peak(s) is higher and that the depletion effect (relative to the second layer) is less pronounced than at a
neutral wall. On the other hand, the density in the depletion zone is lower in the case
of the hydrophilic (than at a neutral) wall. We conclude that, despite its short-range character [see Eq.~(\ref{surf_pref})], the surface field is still effective even at fairly large separations from the wall.

\begin{figure}
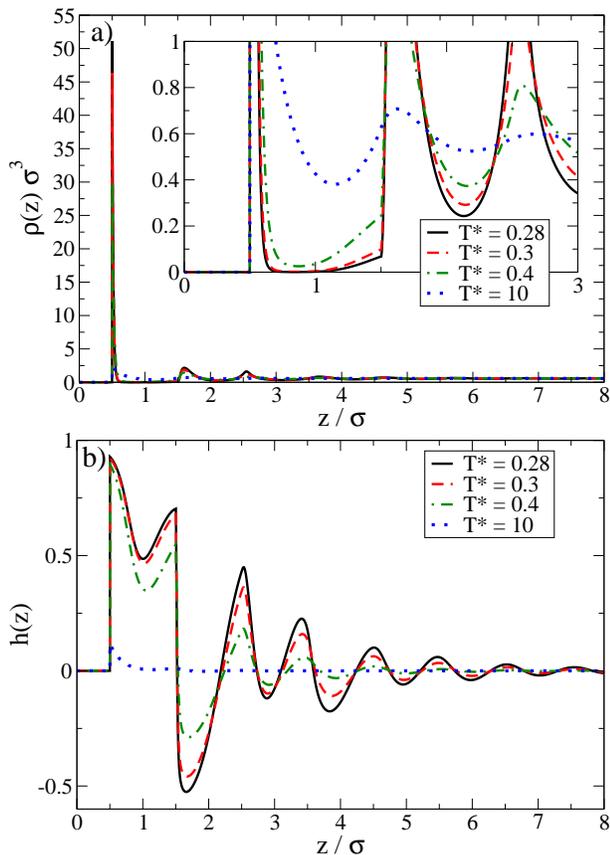

\includegraphics[width=8cm]{Fig6a.eps}
\includegraphics[width=8cm]{Fig6b.eps}
\caption{Same as Fig.~\ref{hydrophilic}, but for a hydrophobic wall characterized by $C_2=-2C$.\label{hydrophobic}} 
\end{figure}

Completely different behavior is observed at a surface preferring the hydrophobic side of the particles. This situation 
is depicted in Fig.~\ref{hydrophobic} where $\eta=0.3$ and we have chosen, for the purpose
of illustration, a fluid-wall coupling parameter twice as large as that of the fluid-fluid interaction [i.e., $C_2=-2C$ in Eq.~(\ref{surf_pref})].
As a result of the dominant surface field, the bilayer formed at neutral walls (see Fig.~\ref{Eta03}) completely breaks down, and one rather observes formation of a monolayer at sufficiently
low temperatures. Moreover, this monolayer is build by particles whose hydrophilic side points {\em away}
 from the wall, contrary to what has been observed before. The subsequent layer then has the reverse orientation and is shifted to slightly
 larger distances ($z\approx 1.65$) as a result of the Janus repulsion between first and second layer. Behind these first two layers, one observes the typical
 oscillatory density and orientation profiles, indicating the tendency for bilayer formation consisting of oppositely oriented Janus particles.
 
Clearly, all of these effects depend on the ratio between the fluid-wall and fluid-fluid coupling parameter, $C_2/C$, 
and on the range parameter $\lambda_2$. Specifically, upon increase of $|C_2/C|$
one observes an increase of the extrema of both, $\rho(z)$ and $h(z)$, and an enhancement of the depletion areas. 
Similar effects emerge upon an increase of $\lambda_2$, that is, an decrease of the range of the surface field.
 \subsection{Confined systems}
 \label{confinement}
\begin{figure}
\includegraphics[width=8cm]{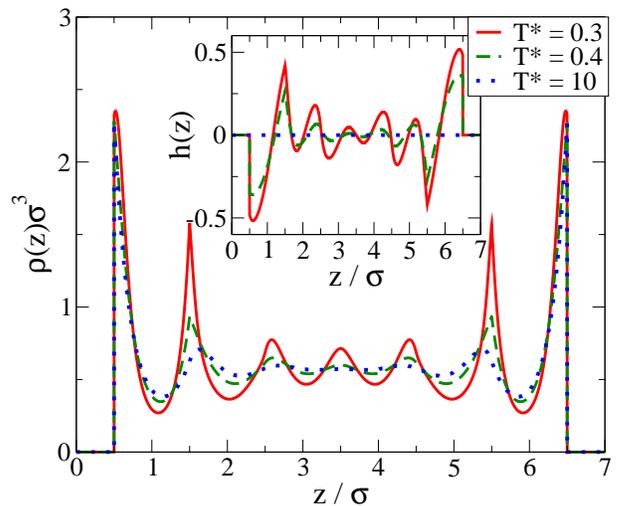}
  \caption{Density profiles and order parameters (inset) at  $\eta=0.3$ and various temperatures in presence of two walls located at $z=0\sigma$ and $z=7\sigma$,
  yielding a separation of $L_{\mathrm{z}}=7\sigma$.\label{Lz_7}} 
  \end{figure}

 Finally, we consider systems confined between {\em two} planar walls, that is, in a slit-pore geometry. For simplicity, we focus
 on the case of neutral walls. For large wall separations $L_{\mathrm{z}}$, one expects the structure
 at either surface to become decoupled from that at the other surface, yielding 
 bulk like-behavior (i.e., $\rho(z)=\rho_{\mathrm{bulk}}$, $h(z)=0$) in between the two walls. Indeed, we have explicitely checked that under such weakly confined conditions, 
 the single-wall behavior discussed in Sec.~\ref{neutral_walls} is recovered at each of the walls. For smaller $L_{\mathrm{z}}$, pronounced confinement effects appear.
 Exemplary density and orientation profiles are plotted in Fig.~\ref{Lz_7}, where $L_{\mathrm{z}}=7\sigma$ (and $\eta=0.3$).
 At high temperatures (i.e., close to the HS limit) the system is almost bulk-like in the center of the slit-pore. Decreasing the temperature we observe layer formation
 (which is typical for any confined system), accompanied by the development of orientational order particularly close to the walls. At the wall separation considered, the ''polarization''
at low temperatures is asymmetric in the sense that particles in the left contact layer point antiparallel to those in the right contact layer. 
This is consistent with the fact that the low-temperature system consists of seven layers, a structure which allows for three full bilayer structures 
(composed of oppositely oriented particles as discussed in Sec.~\ref{neutral_walls}) plus one single layer.

\begin{figure}
\includegraphics[width=8cm]{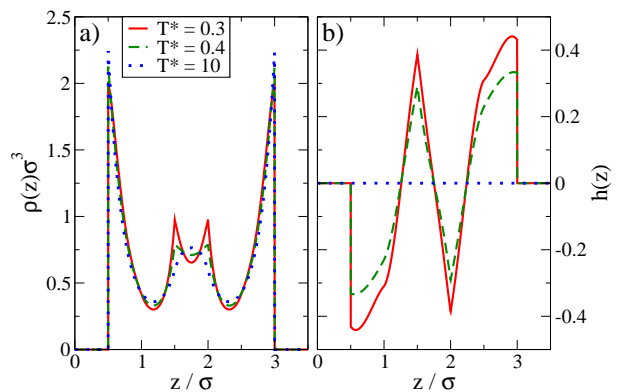}
  \caption{(a) Density profiles and (b) order parameters at  $\eta=0.3$ and various temperatures. The
  walls located at $z=0\sigma$ and $z=3.5\sigma$.\label{Lz_35}} 
  \end{figure}

\begin{figure}
\includegraphics[width=8cm]{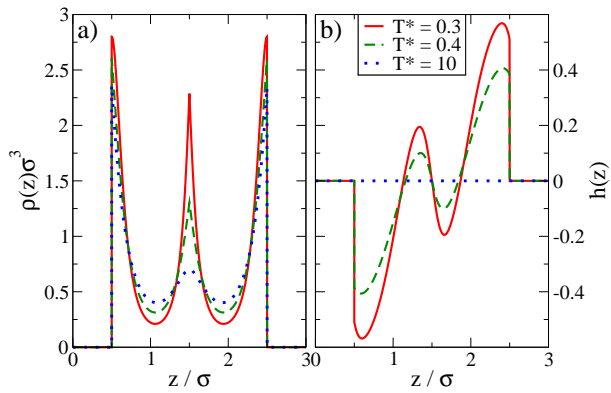}
  \caption{Same as \bild{\ref{Lz_35}}, but for walls located at $z=0\sigma$ and $z=3\sigma$.\label{Lz_3}} 
  \end{figure}

Two more examples are shown in Figs.~\ref{Lz_35} and \ref{Lz_3} corresponding to the cases $L_{\mathrm{z}}=3.5\sigma$ and $L_{\mathrm{z}}=3\sigma$, respectively. 
In the first case,
the high-temperature system is characterized by three layers of particles, with the middle layer being rather thick [see Fig.~\ref{Lz_35}(a)]. Upon lowering the temperature, the anisotropic fluid-fluid interactions yields a splitting of the middle peak, reflecting that particles in the middle layer tend to arrange in a ''buckled'' structure where neighboring particles 
are somewhat shifted to each other with respect to the $z$-direction. The corresponding orientation profile in Fig.~\ref{Lz_35}(b) reveals that, in this buckled middle layer, the particles arrange in an antiparallel way. 
Note that the resulting in-plane arrangement is not particularly unfavorable, since according to our model, side-by-side configurations are energetically ''neutral'' (see Fig.~\ref{Model}).
By assuming this rather complex structure the system at $L_{\mathrm{z}}=3.5\sigma$ overcomes frustration effects. Even stronger confinement, as
it is the case at $L_{\mathrm{z}}=3\sigma$ (see Fig.~\ref{Lz_3}), then yields three pronounced layers of particles,
with the contact layers pointing in opposite direction. However, 
contrary to the situation at $L_{\mathrm{z}}=3.5\sigma$, the order parameter directly at the position of the middle layer is zero. Only slightly left or right of the center
one finds a preferred Janus-like orientation. We therefore regard this case as a frustrated system.

\begin{figure}
\includegraphics[width=8cm]{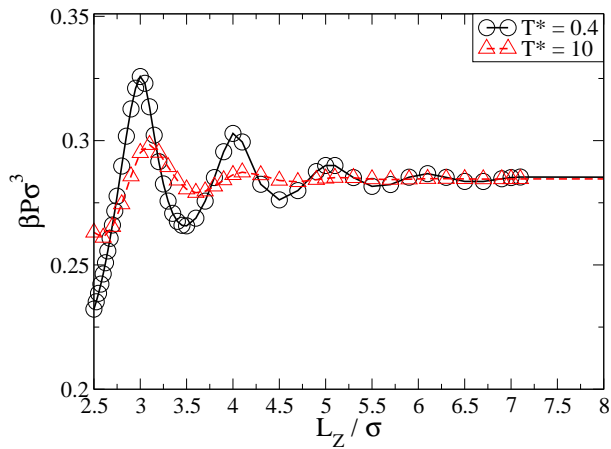}
\caption{Normal pressure as function of the wall separation for two temperatures.\label{pressure_slit}} 
\end{figure}

The different microscopic configurations appearing in the strongly confined systems in dependence of $L_{\mathrm{z}}$
give rise to pronounced oscillations of the normal pressure $P_{\mathrm{z}}$ [as calculated
from the contact theorem, see Eq.~(\ref{contact_theorem_hw})]. The importance of this quantity stems from the fact that it is experimentally accessible, e.g., by colloidal-probe
atomic force microscopy \cite{Klapp08}. Results for $P_{\mathrm{z}}(L_{\mathrm{z}})$ at two temperatures are shown in Fig.~\ref{pressure_slit}.
 As it is typical for confined, dense systems of spherical particles, the oscillations have a period of about one particle diameter. Compared to the high-temperature situation,
 the anisotropic interactions not only strongly enhance
the amplitude of the oscillations; they also lead to slightly asymmetric peak shapes and to a shift of the oscillations. 
Indeed, at low temperatures, the maxima of  $P_{\mathrm{z}}(L_{\mathrm{z}})$ occur at multiples of the particle diameter, 
consistent with the previously discussed frustration effects, e.g., at $L_{\mathrm{z}}=3\sigma$. On the other hand, the complex structure seen, e.g., at $L_{\mathrm{z}}=3.5\sigma$, corresponds
to a minimum of the normal pressure curve.
\section{Summary}
\label{sec:Summary}
In this paper we have used classical density functional theory to explore the structure formation of amphiphilic Janus particles at planar surfaces. 
Our density functional approach involves a sophisticated
(FMT) treatment of the repulsive (HS) interactions, whereas the anisotropic interactions are treated on a mean-field level.
One key finding of our study is 
that, due to the presence of a surface, significant translational and orientational ordering related to bilayer formation 
occurs under conditions where the bulk system is still homogeneous and isotropic. Thus, the
surfaces seem to strongly support the structure formation, even when this surface is just a neutral (hard) wall. 
Moreover, we have shown that the details of the inhomogeneous structure at the wall can be ''tuned'' by varying the surface potential. Indeed, walls preferring the hydrophilic part tend to 
enhance the bilayer structure seen already at neutral walls, whereas hydrophobic walls typically induce a competition between mono- and bilayer structures. 
We note that the degree of hydrophobicity can be experimentally tuned, e.g., by coating silicon wafers with polymer films of varying thickness \cite{Steitz05}.
Finally, we have considered confinement effects emerging from the presence of two planar (neutral) surfaces. It turns out that, for specific wall separations, there are pronounced frustration effects stemming from the interplay between the fluid-surface potential, which prefers planar layering, and the fluid-fluid potential
preferring bilayers with depletion areas in between. This competition is also visible in an experimentally accessible quantity, that is, the normal pressure as function of wall separation.

Clearly, the present study is only a starting point for a more systematic investigation of the impact of surfaces on the self-assembly of amphiphilic Janus particles.
From a physical point of view, one main drawback of our calculations is the restriction to planar, self-assembled geometries.
Especially at low densities one would also expect the occurrence
of spherical structures such as vesicles and micelles; the most ''stable" structure could then be selected by comparison of the related free energies.
Furthermore, in experiments of elongated amphiphilic molecules (rather than amphiphilic spheres) at surfaces, both planar and spherical structures are observed \cite{Dietsch07,Steitz05}, suggesting that different structures could also occur for the spherical (Janus) case.
The strategy to include spherical self-assembled structures within our density functional approach is generally clear, as
shown by Tarazona {\em et al.} \cite{Tarazona} in their study of bulk systems. However, considering such spherical structures in combination with surfaces will require significant additional computational effort due to the further reduction of symmetry. Another open point
is the presence of a vapor-liquid transition in the system. Although the present mean-field approach 
(where the anisotropic contribution to the functional  cancels out in a homogeneous, isotropic state)
does not reveal any condensation transition in the bulk, our preliminary results 
from a modified mean-field approach (see Fig.~\ref{bulk}) indicate that there is, at least, a strong tendency for condensation. 
This is in qualitative agreement to what has been found in another recent study \cite{Pastore} of Janus systems (where, notably, the bulk condensation transition is accompanied
by micellization).  In the context of surface systems,
a bulk condensation transition could have important consequences, since it would enable a wetting transition (on top of the structures already observed). Therefore, it would be very interesting to improve the present density functional approach for the surface systems beyond the mean-field level.
Finally, a further interesting issue concerns the impact of curvature of the substrate on the self-assembled structures \cite{Nonomura,Striolo}.  
Work in these directions is in progress.

\begin{acknowledgments}
  G.~R. would like to thank Professors~R.~Roth, S.~Sokolowski and A.~J.~Archer for fruitful discussions on the numerical implementation
  of FMT. Moreover, he thanks Professor K.~E.~Gubbins for his kind hospitality during his stay at the North Carolina State University in Raleigh, N.C.~Financial support from the DFG via the International Research Training Group "1524 Self-Assembled Soft Matter Nano-Structures at Interfaces" (project B1.1) is gratefully acknowledged. 
  \end{acknowledgments}
%
\appendix
\section{Numerical implementation}
\label{sec:Numericalimplementation}
 In our study, we carry out the minimization numerically, using 
 a one-dimensional lattice with a total length of $L_\mathrm{z}=60\sigma$ and a discretization of $512$-$1024$ points per sphere diameter $\sigma$. We employ
 a simple iterative algorithm. Specifically, the density profile in step $n$ is given by
 \begin{equation}
 \label{mix}
    \rho^{n}\geklammert{z}=\geklammert{1-\alpha}\rho^{n-1}\geklammert{z}+\alpha\rho^{\mathrm{new}}\geklammert{z}\big{|}_{\rho^{n-1}},
 \end{equation}
 where $\alpha$ is a mixing parameter ($0\le\alpha\le 1$) interpolating between the old ($\rho^{n-1}$) and new ($\rho^{\mathrm{new}}$) density profile.
 The new density distribution is calculated via the variational functional derivative
 \begin{equation}
    \rho^{\mathrm{new}}\geklammert{z}\big{|}_{\rho^{n-1}}=\rho_{\mathrm{ideal}}\,\mathrm{exp}\geklammert{-\beta\frac{1}{A}\frac{\delta F_{\mathrm{ex}}[\rho]}{\delta\rho\geklammert{z}}\biggl{|}_{\rho^{n-1}}+\beta\mu_{\mathrm{ex}}}.\label{iterative_formular}
 \end{equation}
 In \formel{\ref{iterative_formular}} $F_{\mathrm{ex}}[\rho]$ is the excess free energy functional
 including both, the HS and the anisotropic, interaction contributions to the full grandcanonical density functional~(\ref{omega}), as well as the contribution of the external potential.
 Furthermore, $\mu_{\mathrm{ex}}$ is the corresponding excess chemical potential, $A$ denotes the surface area,
 and $\rho_{\mathrm{ideal}}$ is the ideal-gas density (of a homogeneous system). There are two constraints for the mixing
 parameter $\alpha$. First, one must ensure in each step that $n_3\geklammert{z}<1$, because of the logarithmic term in \formel{\ref{FMT_1}}. Second, the convergence should be fast enough. 
 In our calculations we used the following strategy (which partially follows that proposed in Ref.~\onlinecite{Roth_FMT}): Within the first iteration steps we calculated the grand potential for several values
 of the mixing parameter. The result was fitted (using a cubic fit) to find the value $\alpha_0$ where the functional becomes minimal. This value $\alpha_0$ was then used
 to define the new density distribution according to Eq.~(\ref{mix}). After few iterations with this procedure (each time updating $\alpha_0$), we kept the value of $\alpha_0$ constant 
 in the further iterations to minimize numerical noise.
 
 As a criterion for convergence we monitored the difference between the old and new density distribution.
 The iteration is stopped when the integral over the absolute values of the difference of both distributions becomes less than $10^{-8}$. This method yields a better accuracy in terms of the boundary conditions, see below.
 
 To calculate the weighted densities, we rewrite \formel{\ref{weighteddensities}} in Fourier space by using the convolution theorem. This yields
 \begin{equation}
      n_\alpha\geklammert{z}=\mathrm{FFT}^{-1}\geklammert{\mathrm{FFT}\geklammert{\rho\geklammert{z}}\cdot\mathrm{FFT}\geklammert{\bar{\omega}_\alpha\geklammert{z}}},\label{FFT_wdensity}
 \end{equation}
 where FFT stands for the fast Fourier transformation \cite{Press}. The analogue of \formel{\ref{FFT_wdensity}} holds for the vectorial weighted densities $\mathbf{n}_\alpha\geklammert{z}$.
 In practice, the weight functions are multiplied with appropriate factors to allow for a polynomial interpolation within the integrals.\\\\
 There are two exact boundary conditions, which give us the opportunity to control the numerical accuracy. First, there is the contact theorem for planar walls \cite{Henderson}
 \begin{equation}
     P=-\int dz \rho\geklammert{z}\int d^2 \omega \alpha\geklammert{z,\bm{\omega}}\frac{\partial}{\partial z}\phi^{\mathrm{surf}}\geklammert{z,\hat{\mathbf{u}}\geklammert{\bm{\omega}}},\label{contact_theorem}
 \end{equation}
  which reduces to
 \begin{equation}
     \rho\geklammert{z=R^+}=\beta P^{\mathrm{bulk}}\label{contact_theorem_hw}
 \end{equation}
 in the case of a hard wall at $z = 0$.
 This theorem expresses the balance of forces, that is, the force per unit area exerted on the wall equals the bulk pressure.
 The second condition is that the density profile far away from the wall approaches the bulk density (in an isotropic system), that is
 \begin{equation}
     \rho\geklammert{z\to\infty}=\rho_{\mathrm{bulk}}.
     \label{rho_limit}
 \end{equation}
Within our calculations, these boundaries conditions are fulfilled with very high accuracy for all systems at neutral walls (see Sec.~\ref{neutral_walls}). In the presence of 
surface fields (see Sec.~\ref{surface_fields}), the contact theorem is fulfilled only with less accuracy, whereas Eq.~(\ref{rho_limit}) still holds.


%

\end{document}